\pdfoutput=1

\documentclass[twocolumn,aps,preprintnumbers,amsmath,amssymb,prl,nofootinbib]{revtex4}

\usepackage{epsfig}
\usepackage{amsmath}
\usepackage{amssymb}
\usepackage{subfigure}
\usepackage{cancel}
\usepackage{textcomp}
\usepackage{calc}
\usepackage{graphicx}
\usepackage{dcolumn}
\usepackage{color}
\usepackage{xcolor}
\usepackage{pifont}

\usepackage{hyperref}
\hypersetup{colorlinks=true,urlcolor=blue,linkcolor=red,citecolor=green!60!black}

\begin{document}


\title{Trans-Planckian Censorship and Inflation in Grand Unified Theories}

\author{Kai Schmitz}
\email{kai.schmitz@cern.ch}
\affiliation{Theoretical Physics Department, CERN, 1211 Geneva 23, Switzerland}
\preprint{CERN-TH-2019-171}


\begin{abstract}

The recently proposed trans-Planckian censorship conjecture (TCC) amounts to the claim that inflation models with an inflationary energy scale larger than $\Lambda_{\rm inf}^{\rm max} \sim 10^9\,\textrm{GeV}$ belong to the swampland, \textit{i.e.}, cannot be  embedded into a consistent theory of quantum gravity.
In this paper, we point out that this constraint can be readily satisfied in D-term hybrid inflation (DHI), which is a well-motivated inflation scenario in the context of supersymmetric grand unification.
In DHI, the amplitude of the primordial scalar power spectrum originates from a Fayet--Iliopoulos term of the order of the unification scale, $\sqrt{\xi} \sim 10^{16}\,\textrm{GeV}$.
At the same time, the TCC results in an upper bound on the corresponding gauge coupling constant of $g_{\rm max} \sim 10^{-14}$.
We are able to show that this constraint translates into an upper bound on the gravitino mass of $m_{3/2}^{\rm max} \sim 10\,\textrm{MeV}$, which opens the possibility that dark matter is accounted for by thermally produced gravitinos, if the reheating temperature is close to $T_{\rm reh} \sim 100\,\textrm{TeV}$.
Interestingly enough, a somewhat similar gravitino mass range has recently been derived in a model that aims at explaining dark energy in terms of axion quintessence and resolving the Hubble tension by means of decaying gravitino dark matter.
\end{abstract}


\date{\today}
\maketitle


\noindent\textbf{Introduction.}
A standard assumption in modern cosmology is that the initial conditions of the Hot Big Bang were determined by cosmic inflation, a stage of accelerated expansion in the early Universe~\cite{Starobinsky:1980te,Guth:1980zm,Sato:1980yn,Linde:1981mu,Albrecht:1982wi}.
Inflation was originally conceived to explain the vast size of the observable Universe as well as its high degree of homogeneity and isotropy on cosmological scales; but in addition, inflation can also provide the seeds for the postinflationary formation of structure on smaller scales (see~\cite{Linde:2005ht,Lyth:1998xn} for reviews).
One of its key properties is that the spatial extent of causally connected Hubble patches remains roughly constant during inflation, while at the same time, microscopic quantum fluctuations are stretched to macroscopic size.
This means that primordial (scalar and tensor) perturbation modes continuously exit the Hubble horizon during inflation, whereupon they ``freeze out'' and classicalize.
During the subsequent decelerated expansion, these classical perturbations then re-enter the Hubble horizon, which causes them to ``thaw'' and evolve further.
The primordial scalar perturbations thus lead to the temperature anisotropies in the cosmic microwave background (CMB)~\cite{Aghanim:2018eyx,Akrami:2018odb} and eventually to the large-scale structure of the Universe, while the primordial tensor perturbations give rise to a (yet unobserved) stochastic background of gravitational waves~\cite{Guzzetti:2016mkm} (see also~\cite{DEramo:2019tit}).


Physical length scales are exponentially stretched during inflation.
This characteristic feature is essential in explaining how it is possible that the entire observable Universe was initially contained in a single causal patch.
At the same time, it implies that sub-Planckian length scales, $\ell < \ell_{\rm Pl} = M_{\rm Pl}^{-1}$ (where $M_{\rm Pl} \simeq 2.435 \times 10^{18}\,\textrm{GeV}$ denotes the reduced Planck mass) can be easily stretched to super-horizon size if inflation lasts long enough.
This is known as the \textit{trans-Planckian problem} of inflationary cosmology~\cite{Martin:2000xs} (see also~\cite{Brandenberger:2000wr,Niemeyer:2000eh,Kempf:2001fa,Easther:2001fi,Bozza:2003pr,Brandenberger:2012aj}).
Indeed, regarding general relativity (GR) as the low-energy effective field theory (EFT) of some unknown theory of quantum gravity, one expects perturbative calculations in GR to remain valid only up to some cut-off scale $\Lambda_{\rm EFT} \sim M_{\rm Pl}$.
For modes on sub-Planckian length scales, the standard approach of calculating the inflationary power spectrum therefore ventures beyond the validity range of the EFT.
In line with this observation, Bedroya and Vafa recently conjectured that such a situation may never occur in a consistent theory of quantum gravity~\cite{Bedroya:2019snp}.
That is, according to the \textit{trans-Planckian censorship conjecture} (TCC), any inflation model that stretches modes on sub-Planckian length scales to the extent that they eventually exit the Hubble horizon must lie in the swampland~\cite{Vafa:2005ui}.
The TCC is thus the newest member in the family of quantum gravity swampland conjectures~\cite{ArkaniHamed:2006dz,Ooguri:2006in,Obied:2018sgi,Agrawal:2018own,Garg:2018reu,Ooguri:2018wrx} (see~\cite{Palti:2019pca} for a review).


The TCC has important implications for inflationary cosmology~\cite{Bedroya:2019tba} (see also~\cite{Cai:2019hge,Tenkanen:2019wsd,Das:2019hto,Mizuno:2019bxy,Brahma:2019unn,Dhuria:2019oyf,Torabian:2019zms,Cai:2019igo,Kadota:2019dol}).
In particular, it severely constrains the allowed range for the total number of $e$-folds $N_{\rm inf}$ during inflation, which translates into upper bounds on the inflationary  Hubble rate $H_{\rm inf}$, energy scale $\Lambda_{\rm inf} = V_{\rm inf}^{1/4}$ (where $V_{\rm inf}$ is the inflaton potential), and tensor-to-scalar ratio $r$.
The main purpose of this paper is to demonstrate that these bounds can be readily satisfied in D-term hybrid inflation (DHI)~\cite{Binetruy:1996xj,Halyo:1996pp}, which is a well-motivated inflation scenario in the context of supersymmetric grand unified theories (GUTs).
A remarkable outcome of our analysis is that the dimensionful input parameter of DHI, namely, the Fayet--Iliopoulos (FI) term~\cite{Fayet:1974jb} can still be chosen to be of the order of GUT scale, $\sqrt{\xi} \sim 10^{16}\,\textrm{GeV}$, despite the severe TCC bound on the inflationary energy scale.
This hierarchy of scales can be achieved in a technically natural way by choosing a tiny gauge coupling constant, $g \lesssim 10^{-14}$.
At the same time, the best-fit value for the scalar spectral index, $n_s^{\rm obs} \simeq 0.9649$~\cite{Aghanim:2018eyx,Akrami:2018odb}, can be reproduced for gravitino masses in the sub-GeV range.
This points toward gravitino dark matter (DM) and the mediation of supersymmetry (SUSY) breaking via gauge mediation.


\noindent\textbf{Trans-Planckian censorship conjecture.}
We begin by reviewing the TCC bounds on inflation.
According to the TCC, physical length scales that are initially Planckian must not exit the Hubble radius $H^{-1}$ during inflation,
\begin{align}
\label{eq:tcc}
e^{N_{\rm inf}}\,\ell_{\rm Pl} < H_{\rm end}^{-1} \,.
\end{align}
$N_{\rm inf} = \ln\left(a_{\rm end}/a_{\rm ini}\right)$ is the total number of $e$-folds during inflation (where $a_{\rm ini}$ and $a_{\rm end}$ denote the Friedmann-Lema\^itre-Robertson-Walker scale factor at the beginning and at the end of inflation, respectively), and $H_{\rm end}^{-1}$ is the Hubble radius at the end of inflation.
Eq.~\eqref{eq:tcc} results in an upper bound on the number of $e$-folds, $N_{\rm inf} < N_{\rm inf}^{\rm max}$,
\begin{align}
N_{\rm inf}^{\rm max} = \ln\left(\frac{M_{\rm Pl}}{H_{\rm end}}\right) \simeq 43.5 + \ln\left(\frac{0.3\,\textrm{GeV}}{H_{\rm end}}\right) \,,
\end{align}
where we anticipated an exceptionally small $H_{\rm end}$.


Meanwhile, inflation is also required to solve the horizon and flatness problems of the Hot Big Bang, which means that the largest observable scale today, $\ell_0$, must have initially been contained in a single causal volume.
Let us denote the number of $e$-folds before the end of inflation when the corresponding comoving scale crossed outside the horizon by $N_{\ell_0}$. 
We are then able to write
\begin{align}
\ell_0 = \frac{a_0}{a_{\rm reh}}\frac{a_{\rm reh}}{a_{\rm end}}\,e^{N_{\ell_0}} H_{\ell_0}^{-1} \,,
\end{align}
where $H_{\ell_0}^{-1}$ is the Hubble radius at the time of horizon crossing, and $a_{\rm reh}$ and $a_0$ denote the scale factor at the end of reheating and at the present time, respectively.
The ratio $a_{\rm reh}/a_{\rm end}$ depends on the energy density at the end of inflation, $\rho_{\rm end} = 3\,M_{\rm Pl}^2H_{\rm end}^2$, the energy density at the end of reheating, $\rho_{\rm reh} = \pi^2/30\,g_{*,\rho}^{\rm reh}\,T_{\rm reh}^4$, and the equation-of-state parameter $w = p/\rho$ during reheating,
\begin{align}
\label{eq:Nreh}
N_{\rm reh} = \ln\left(\frac{a_{\rm reh}}{a_{\rm end}}\right) = \frac{1}{3\left(1+w\right)}\ln\left(\frac{\rho_{\rm end}}{\rho_{\rm reh}}\right) \,.
\end{align}
The ratio $a_0/a_{\rm reh}$, on the other hand, can be determined if one assumes entropy to be conserved after reheating,
\begin{align}
\label{eq:N0}
N_0 = \ln\left(\frac{a_0}{a_{\rm reh}}\right) = \frac{1}{3}\ln\bigg(\frac{g_{*,s}^{\rm reh}}{g_{*,s}^0}\bigg) + \ln\left(\frac{T_{\rm reh}}{T_0}\right)\,.
\end{align}
Below, we will treat the Hubble rate $H_{\rm end}$ and the reheating temperature $T_{\rm reh}$ as free parameters, while fixing all other parameters in Eqs.~\eqref{eq:Nreh} and \eqref{eq:N0} at specific values.
We will restrict ourselves to the standard case of matter-dominated expansion during reheating, $w = 0$;
the CMB photon temperature at the present time is given by $T_0 \simeq 2.725\,\textrm{K}\simeq 2.349 \times 10^{-13}\,\textrm{GeV}$~\cite{Fixsen:2009ug}; 
and for the effective numbers of relativistic degrees of freedom, we will use the values in the minimal supersymmetric standard model (MSSM), $g_{*,\rho}^{\rm reh} = g_{*,s}^{\rm reh} = 915/4$ and $ g_{*,s}^0 = 43/11$.


In any viable model of inflation, $N_{\ell_0}$ must be smaller than the maximal number of $e$-folds $N_{\rm inf}^{\rm max}$.
Combining all of the expressions above, this requirement, $N_{\ell_0} < N_{\rm inf}^{\rm max}$, can be used to derive an upper bound on the Hubble rate,
\begin{widetext}
\begin{align}
\label{eq:Hbound}
H_{\rm inf} < H_{\rm inf}^{\rm max} = M_{\rm Pl}\left[\bigg(\frac{g_{*,s}^{\rm reh}}{g_{*,s}^0}\bigg)^{1/3}\left(\frac{90}{\pi^2 g_{*,\rho}^{\rm reh}}\right)^{\alpha/2}\left(\frac{M_{\rm Pl}}{T_{\rm reh}}\right)^{2\alpha-1}\frac{1}{T_0\,\ell_0}\right]^{1/\left(2-\alpha\right)} \,, \qquad \alpha = \frac{2}{3\left(1+w\right)} \,,
\end{align}
\end{widetext}
where $\alpha$ characterizes the time dependence of the scale factor during reheating, $a\left(t\right) \propto t^\alpha$.
Formally, the Hubble rate $H_{\rm inf}$ in Eq.~\eqref{eq:Hbound} is defined as $H_{\rm inf} = H_{\ell_0}^\beta H_{\rm end}^{1-\beta}$, where $\beta = 1/\left(2-\alpha\right)$.
In this sense, the bound in Eq.~\eqref{eq:Hbound} is completely general.%
\footnote{In particular, it is more general than the bounds in~\cite{Bedroya:2019tba,Tenkanen:2019wsd,Mizuno:2019bxy}, which are based on additional assumptions, such as instantaneous reheating or a quasi-constant Hubble rate during inflation.}
In practice, however, one can work with $H_{\rm inf} \approx H_{\ell_0} \approx H_{\rm end}$, as any inflation model consistent with the bound in Eq.~\eqref{eq:Hbound} anyway results in a small variation of the Hubble rate during inflation, $-\dot{H}/H^2 \ll 1$.


In passing, we also mention that Eq.~\eqref{eq:Hbound} can be generalized to account for the possibility of entropy production after reheating.
Suppose the comoving entropy density changes by a factor $\gamma \geq 1$ between reheating and today,
\begin{align}
\gamma = \frac{g_{*,s}^0T_0^3a_0^3}{g_{*,s}^{\rm reh}T_{\rm reh}^3a_{\rm reh}^3} \,.
\end{align}
Late-time entropy production can thus be described by rescaling $g_{*,s}^{\rm reh}$ by one power of the factor $\gamma$~\cite{Ji:2019gfy}, such that
\begin{align}
H_{\rm inf}^{\rm max} \rightarrow \gamma^{1/3/\left(2-\alpha\right)} H_{\rm inf}^{\rm max} \,.
\end{align}
This rescaling allows one to weaken the bound on $H_{\rm inf}$ by considering a large amount of entropy production after reheating, $\gamma \gg 1$.
In the remainder of this paper, we will, however, ignore this option and focus on the $\gamma = 1$ case.


To evaluate the right-hand side of Eq.~\eqref{eq:Hbound}, one needs to specify the scale $\ell_0$. 
The authors of~\cite{Bedroya:2019tba,Tenkanen:2019wsd,Dhuria:2019oyf} identified $\ell_0$ with the present Hubble radius, $H_0^{-1} \simeq 3.0/h\,\textrm{Gpc}$ (where $H_0 = 100\,h\,\textrm{km}/\textrm{s}/\textrm{Mpc}$ and $h_{\rm cmb} \simeq 0.6736$ according to the 2018 PLANCK data~\cite{Aghanim:2018eyx,Akrami:2018odb}).
Meanwhile, the authors of~\cite{Mizuno:2019bxy,Torabian:2019zms} worked with the comoving scale that was of horizon size at the time of matter-dark-energy equality, $a_0/k_\Lambda = 2^{-1/2}\,\Omega_\Lambda^{-1/6}\,\Omega_{\rm m}^{-1/3} H_0^{-1} \simeq 3.3/h\,\textrm{Gpc}$, which is very close to the largest comoving scale that ever became horizon-sized during the decelerated expansion, $a_0/k_{\rm min} = 2^{1/3}\,3^{-1/2}\,\Omega_\Lambda^{-1/6}\,\Omega_{\rm m}^{-1/3} H_0^{-1} \simeq 3.4/h\,\textrm{Gpc}$.
In our analysis, we will consider yet another scale\,---\,the optical horizon $d_{\rm opt}\simeq 9.4/h\,\textrm{Gpc}$ of our Universe~\cite{Mukhanov:2005sc},
\begin{align}
\label{eq:dopt}
\frac{d_{\rm opt}}{H_0^{-1}} = \int_{a_{\rm dec}}^{a_0} \frac{da}{a_0}\left[\Omega_{\rm r} + \frac{a}{a_0}\,\Omega_{\rm m} + \left(\frac{a}{a_0}\right)^4\Omega_\Lambda\right]^{-1/2} \,,
\end{align}
where $a_{\rm dec}$ denotes the scale factor at CMB decoupling.
Note that $d_{\rm opt}$ is almost identical to our particle horizon, which follows from Eq.~\eqref{eq:dopt} by taking the limit $a_{\rm dec} \rightarrow 0$.


Upon identifying $\ell_0$ with $d_{\rm opt}$, Eq.~\eqref{eq:Hbound} results in
\begin{align}
\label{eq:Hmax}
H_{\rm inf}^{\rm max} \simeq 0.35\,\textrm{GeV} \left(\frac{h}{h_{\rm cmb}}\right)^{3/4}\left(\frac{100\,\textrm{TeV}}{T_{\rm reh}}\right)^{1/4} \,,
\end{align}
where we now fixed all other parameters as described above.
This constraint immediately translates into upper bounds on the inflationary energy scale, $\Lambda_{\rm inf} < \Lambda_{\rm inf}^{\rm max}$,
\begin{align}
\label{eq:Lmax}\hspace{-0.1cm}
\Lambda_{\rm inf}^{\rm max} \simeq 1.2 \times 10^9\,\textrm{GeV} \left(\frac{h}{h_{\rm cmb}}\right)^{3/8}\left(\frac{100\,\textrm{TeV}}{T_{\rm reh}}\right)^{1/8} \,,
\end{align}
and the tensor-to-scalar ratio, $r = A_t/A_s < r_{\rm max}$,
\begin{align}
\label{eq:rmax}
r_{\rm max} \simeq 2.0 \times 10^{-30} \left(\frac{h}{h_{\rm cmb}}\right)^{3/2}\left(\frac{100\,\textrm{TeV}}{T_{\rm reh}}\right)^{1/2} \,.
\end{align}
Here, we used the measured value of the amplitude of the scalar power spectrum, $A_s^{\rm obs} \simeq 2.10\times10^{-9}$~\cite{Aghanim:2018eyx,Akrami:2018odb}, as well as two standard relations in the description of slow-roll inflation, $V_{\rm inf} = 3\,M_{\rm Pl}^2H_{\rm inf}^2$ and $A_t = 2\,(H_{\rm inf}/(\pi M_{\rm Pl}))^2$.


\medskip\noindent\textbf{D-term hybrid inflation.}
The bounds in Eqs.~\eqref{eq:Hmax}, \eqref{eq:Lmax}, and \eqref{eq:rmax} represent severe constraints on the dynamics of cosmic inflation.
Nonetheless, it is straightforward to satisfy these constraints in DHI, as we will now discuss.
First of all, we recall that supersymmetric hybrid inflation is an attractive model of inflation on general grounds because it allows one to establish a connection between cosmology (inflation) and particle physics (grand unification).
The crucial observation is that hybrid inflation ends via a rapid second-order phase transition~\cite{Linde:1991km,Linde:1993cn}.
This so-called waterfall transition can be identified as the spontaneous breaking of (a subgroup of) a local GUT symmetry, such as, \textit{e.g.}, $U\left(1\right)_{B-L}$~\cite{Buchmuller:2010yy,Buchmuller:2011mw,Buchmuller:2012wn,Buchmuller:2012bt,Buchmuller:2013lra,Schmitz:2012kaa,Buchmuller:2013dja,Domcke:2017xvu,Domcke:2017rzu}.


In this paper, we will not specify an explicit GUT embedding of DHI but simply focus on the minimal scenario based on a $U(1)$ FI term.
Similarly, we will also refrain from considering F-term hybrid inflation (FHI)~\cite{Copeland:1994vg,Dvali:1994ms} as an alternative to DHI.
This model comes with less parametric freedom than DHI (unlike in DHI, the $U(1)$ gauge coupling constant $g$ plays no role in FHI) and requires a more complicated treatment due to the fact that it is a two-field model of inflation~\cite{Buchmuller:2014epa} (see also~\cite{Rehman:2009nq}).
A detailed discussion of both FHI and DHI can be found in~\cite{Schmitz:2018nhb}.
Let us now review some of the results on DHI derived in~\cite{Schmitz:2018nhb} that will allow us to satisfy the TCC bound in Eq.~\eqref{eq:Hbound}.


As in~\cite{Schmitz:2018nhb}, we will consider DHI in combination with a SUSY-breaking Polonyi sector~\cite{Polonyi:1977pj} (see also \cite{Buchmuller:2000zm}).
The superpotential of our model is therefore given as follows,
\begin{align}
\label{eq:W}
W = \kappa\,S\,\Phi\bar{\Phi} + \mu_X^2\,X + W_0 \,.
\end{align}
Here, $S$ denotes the inflaton field, $\Phi$ and $\bar{\Phi}$ are two oppositely charged waterfall fields, and $X$ is the Polonyi field.
$\kappa$ represents the dimensionless inflaton Yukawa coupling, $\mu_X$ is the SUSY-breaking scale in the Polonyi sector, and $W_0$ is an $R$-symmetry-breaking constant that is fixed by the requirement that inflation must end in a Minkowski vacuum with zero cosmological constant.
We assume that the Polonyi field is stabilized at $\left<X\right> = 0$  in consequence of additional hidden-sector interactions.
$\mu_X$ and $W_0$ can thus be related to the gravitino mass $m_{3/2}$ as follows,
\begin{align}
\mu_X^2 = \sqrt{3}\,M_{\rm Pl}\,m_{3/2} \,,\quad W_0 = m_{3/2}\,M_{\rm Pl}^2 \,.
\end{align}
We will work with a canonical K\"ahler potential for all fields in Eq.~\eqref{eq:W},
supplemented by a Planck-suppressed coupling between the inflaton and the Polonyi field,
\begin{align}
K \supset \frac{\chi}{M_{\rm Pl}^2}\,S^\dagger S\, X^\dagger X \,.
\end{align}
This operator is allowed by all symmetries and expected to be present in the EFT at energies below $\Lambda_{\rm EFT} \sim M_{\rm Pl}$.
Below, we will set the $\mathcal{O}\left(1\right)$ Wilson coefficient $\chi$ to unity, for simplicity (see also \cite{Schmitz:2018nhb}).
Finally, we assume a nonvanishing FI term in the D-term scalar potential,%
\footnote{Constant FI terms cannot be consistently coupled to supergravity~\cite{Komargodski:2009pc,Dienes:2009td} (see also \cite{Cribiori:2017laj,Kuzenko:2018jlz,Antoniadis:2018cpq}).
We therefore assume an effective field-dependent FI term, \textit{i.e.}, an FI term that is generated in consequence of spontaneous symmetry breaking and whose magnitude is controlled by the vacuum expectation values (VEVs) of a number of moduli, $\xi = \xi\left(\left<\varphi_i\right>\right)$ (see~\cite{Domcke:2014zqa} for an explicit model).
\label{fn:xi}}
\begin{align}
\label{eq:VD}
V_D = \frac{g^2}{2}\left[\xi - \left(\left|\phi\right|^2 - \left|\bar{\phi}\right|^2\right)\right]^2 \,,
\end{align}
which is responsible for the nonvanishing vacuum energy density that drives inflation, $V_D^0 = 3\,M_{\rm Pl}^2H_{\rm inf}^2 = 1/2\,g^2\xi^2$.


Eqs.~\eqref{eq:W} to \eqref{eq:VD} fully specify our inflaton model.
In particular, they enable us to compute the scalar potential for the real inflaton field, $s = \sqrt{2}\left|S\right|$.
In this potential, inflation takes place at field values larger than the critical field value $s_{\rm crit} = g/\kappa\,v$, which marks the location of the tachyonic instability in the potential that triggers the waterfall phase transition.
Here, $v = \sqrt{2}\left<\Phi\right> = \sqrt{2\xi}$ denotes the VEV of the symmetry-breaking waterfall field after the phase transition.
In the following, we will consider inflaton Yukawa couplings $\kappa$ of $\mathcal{O}\left(0.1 \cdots 1\right)$.
As discussed in~\cite{Schmitz:2018nhb}, this implies that the CMB pivot scale $\ell_* = 20\,\textrm{Mpc}$ exits the Hubble horizon at a field value far away from the instability, $s_* \gg s_{\rm crit}$.
In addition, we will assume a tiny $U(1)$ gauge coupling constant, $g \ll \kappa$, in order to suppress the inflationary Hubble rate.%
\footnote{The opposite scenario, $\kappa \ll g$, has been considered in~\cite{Buchmuller:2014rfa,Buchmuller:2014dda} (see also~\cite{Bryant:2016tzg,Bryant:2016sjj,Ishiwata:2018dxg,Gunji:2019wtk}), which results in a super-Planckian critical field value, $s_{\rm crit} > M_{\rm Pl}$.
In this case, inflation occurs during the waterfall transition and leads to an $\mathcal{O}\left(0.1\right)$ tensor-to-scalar ratio.\smallskip}
In this parameter regime, we obtain the following scalar potential,
\begin{align}
\label{eq:V}
V \simeq V_D^0 + \frac{1}{2}\,m_s^2 \,s^2 + V_{1\ell}^0\,\ln\left(\frac{s}{s_{\rm crit}}\right) \,.
\end{align}
The mass parameter $m_s$ stems from soft SUSY breaking,
\begin{align}
m_s^2 = \left(1-3\,\chi\right) m_{3/2}^2 \,,
\end{align}
while the logarithmic term arises in consequence of one-loop radiative corrections when integrating out the heavy waterfall fields $\Phi$ and $\bar{\Phi}$ along the inflationary trajectory,
\begin{align}
V_{1\ell}^0 = \frac{m_D^4}{8\pi^2} \,,\quad m_D = g\sqrt{\xi} \,.
\end{align}
Note that, in consequence of our choice of $\chi$, the inflaton mass turns out to be tachyonic during inflation, $m_s^2 < 0$.
If this was not the case, \textit{i.e.}, for $\chi< 1/3$, we would not be able to reproduce the best-fit value of the spectral index.


Based on the potential in Eq.~\eqref{eq:V}, one is now able to perform a standard analysis of slow-roll inflation.
We begin by writing down the slow-roll equation of motion,
\begin{align}
\label{eq:eom}
\frac{ds^2}{dN} = 2\,\Delta\left(s^2 - s_{\rm max}^2\right) \,,
\end{align}
where the quantities $\Delta$ and $s_{\rm max}$ are defined as follows,
\begin{align}
\Delta = \frac{m_s^2 M_{\rm Pl}^2}{V_D^0} \,,\quad s_{\rm max} = \left|\frac{V_{1\ell}^0}{m_s^2}\right|^{1/2} \,.
\end{align}
$s_{\rm max}$ marks the position of a hilltop (\textit{i.e.}, local maximum) in the potential.
Slow-roll inflation takes place at field values below $s_{\rm max}$ and above $s_{\rm fast}$, where the latter represents the field value at which the slow-roll condition $\eta = M_{\rm Pl}^2\,V''/V < \eta_{\rm max} = 10^{-0.5}$ becomes violated,%
\footnote{For $\kappa \lesssim 10^{-3}$, the hilltop moves in the direct vicinity of the tachyonic instability, $s_{\rm max} \simeq s_{\rm crit}$, such that successful inflaton can only be realized in a very small region in field space~\cite{Schmitz:2018nhb}.
This problem is absent for $\kappa \sim 0.1 \cdots 1$, which leads to $s_{\rm max} \gg s_{\rm crit}$.}
\begin{align}
s_{\rm fast} = \left|\frac{\Delta}{\eta_{\rm max} + \Delta}\right|^{1/2}\,s_{\rm max} \,.
\end{align}
Imposing the boundary condition $s = s_{\rm fast}$ at $N = 0$, the slow-roll equation in Eq.~\eqref{eq:eom} has the following solution,
\begin{align}
s^2\left(N\right) = s_{\rm max}^2\left[1 + F\left(N\right)\right] \,,
\end{align}
with $F$ as a function of the number of $e$-folds $N$ given as 
\begin{align}
F\left(N\right) = - \left(1 + \frac{\Delta}{\eta_{\rm max} + \Delta}\right) e^{2N\Delta} \,.
\end{align}
This solution for $s\left(N\right)$ allows one to work out the slow-roll parameters $\varepsilon = 1/2\,M_{\rm Pl}^2\left(V'/V\right)^2$ and $\eta = M_{\rm Pl}^2\,V''/V$,
\begin{align}
\varepsilon = \left(\frac{s_{\rm max}}{M_{\rm Pl}}\right)^2\frac{\left(F\Delta\right)^2}{2\left(1+F\right)} \,,\quad \eta = \frac{\left(2+F\right)\Delta}{1+F} \,,
\end{align}
which in turn can be used to compute the amplitude,
\begin{align}
\label{eq:As}
A_s = \left.\frac{V}{24\pi^2\,\varepsilon\,M_{\rm Pl}^4}\vphantom{\left(\frac{\sqrt{\xi}}{M_{\rm Pl}}\right)^4}\right|_{N_*} = \left.\frac{1+F}{6\left|\Delta\right|F^2}\left(\frac{\sqrt{\xi}}{M_{\rm Pl}}\right)^4\right|_{N_*} \,,
\end{align}
and index of the primordial scalar power spectrum,
\begin{align}
\label{eq:ns}
n_s \approx \left.1 + 2\eta\vphantom{\frac{2\left(2+F\right)\Delta}{1+F}}\right|_{N_*} = \left.1 + \frac{2\left(2+F\right)\Delta}{1+F}\right|_{N_*} \,.
\end{align}
Here, we neglected $\varepsilon \ll \eta$ in computing $n_s = 1 + 2\,\eta - 6\,\varepsilon$.
All $N$-dependent quantities in Eqs.~\eqref{eq:As} and \eqref{eq:ns} need to be evaluated at $N_*$, \textit{i.e.}, the number of $e$-folds between the time of CMB horizon exit and the end of inflation,
\begin{align}
N_* & = \ln\left(\ell_* H_{\rm inf}\right) - N_0 - N_{\rm reh}
\\ \nonumber
& \simeq 36.8 + \frac{1}{3}\ln\left(\frac{H_{\rm inf}}{0.3\,\textrm{GeV}}\right) + \frac{1}{3}\ln\left(\frac{T_{\rm reh}}{100\,\textrm{TeV}}\right) \,.
\end{align}


Remarkably enough, the expressions in Eqs.~\eqref{eq:As} and \eqref{eq:ns} offer enough parametric freedom to simultaneously satisfy the two constraints $A_s = A_s^{\rm obs} \simeq 2.10\times10^{-9}$ and $n_s = n_s^{\rm obs} \simeq 0.9649$.
For $\eta_{\rm max} = 10^{-0.5}$ and $N_* = 36.8$, the condition $n_s = n_s^{\rm obs}$ results in a unique value for $\Delta$,
\begin{align}
\Delta = -\frac{2}{3}\left(\frac{m_{3/2}}{H_{\rm inf}}\right)^2 \simeq -3.0 \times 10^{-3} \,.
\end{align}
With $\Delta$ fixed at this value, $A_s$ becomes a function of the FI parameter $\xi$ only.
$A_s = A_s^{\rm obs}$ is then satisfied for
\begin{align}
\sqrt{\xi} \simeq 0.80 \times 10^{16}\,\textrm{GeV} \,,
\end{align}
or $v \simeq 1.1 \times 10^{16}\,\textrm{GeV}$ in terms of the VEV of the waterfall field $\Phi$.
This essentially coincides with the value of the GUT scale that one expects in SUSY GUT models, thus underlining the close connection between DHI and supersymmetric grand unification.
With $\Delta$ determined by $n_s^{\rm obs}$ and $\sqrt{\xi}$ by $A_s^{\rm obs}$, the Hubble rate $H_{\rm inf}$ is now solely controlled by the gauge coupling constant $g$,
\begin{align}
\label{eq:Hinfg}
H_{\rm inf} & \simeq 0.11\,\textrm{GeV}\left(\frac{g}{10^{-14}}\right) \,,
\end{align}
and similarly for the gravitino mass $m_{3/2}$,
\begin{align}
\label{eq:m32g}
m_{3/2} & \simeq 7.2\,\textrm{MeV}\left(\frac{g}{10^{-14}}\right) \,.
\end{align}
Eqs.~\eqref{eq:Hinfg} and \eqref{eq:m32g} are the main technical results of this paper.
We observe that, by choosing a very small gauge coupling constant $g$, DHI always allows one to satisfy the TCC bounds on inflation, while at the same time being in accord with the best-fit values for $A_s^{\rm obs}$ and $n_s^{\rm obs}$.
Indeed, by comparing $H_{\rm inf}$ in Eq.~\eqref{eq:Hinfg} with the maximal value $H_{\rm inf}^{\rm max}$ in Eq.~\eqref{eq:Hmax}, we obtain an upper bound on $g$,
\begin{align}
g_{\rm max} \simeq 3.3\times 10^{-14}\left(\frac{h}{h_{\rm cmb}}\right)^{3/4}\left(\frac{100\,\textrm{TeV}}{T_{\rm reh}}\right)^{1/4} \,,
\end{align}
which immediately translates into a maximal $m_{3/2}$ value, 
\begin{align}
\label{eq:m32max}
m_{3/2}^{\rm max} \simeq 23\,\textrm{MeV}\left(\frac{h}{h_{\rm cmb}}\right)^{3/4}\left(\frac{100\,\textrm{TeV}}{T_{\rm reh}}\right)^{1/4} \,.
\end{align}


Our results in this section serve as a proof of principle that a small gauge coupling constant in a GUT-inspired scenario can lead to an inflationary energy scale that is parametrically smaller than the GUT scale and thus consistent with the TCC bound in Eq.~\eqref{eq:Hbound}.
Of course, our model does not address the question why $g$ should be so small, nor does it explain why the supergravity and radiative corrections should conspire in such a way that the scalar potential ends up being extremely flat.
These questions are beyond the scope of this work and need to be addressed in the UV completion of our model (see, \textit{e.g.},~\cite{Halyo:1999fd,Halyo:1999wn} for related earlier work in string theory).


\medskip\noindent\textbf{Discussion.}
In light of the results derived in the previous section, several comments are in order.


\smallskip\noindent\textit{Distance and de Sitter conjectures.}
First of all, we point out that DHI also trivially satisfies the swampland distance conjecture~\cite{Ooguri:2006in}, which claims that scalar fields can at most traverse an $\mathcal{O}\left(M_{\rm Pl}\right)$ distance in field space.
In our case, the inflaton field range is always bounded from above by the local maximum at $s_{\rm max}$, by construction,
\begin{align}
\label{eq:smax}
s_{\rm max} = \frac{g\,M_{\rm Pl}}{2\pi\sqrt{\left|\Delta\right|}} \simeq 7.1\times10^4\,\textrm{GeV} \left(\frac{g}{10^{-14}\,\textrm{GeV}}\right) \,,
\end{align}
which is parametrically smaller than the Planck scale.
The small gauge coupling constant $g$ is therefore instrumental in satisfying both the TCC and the distance conjecture.
At the same time, there is a slight tension between DHI and the refined de Sitter conjecture~\cite{Garg:2018reu,Ooguri:2018wrx}, just as in most other models of single-field slow-roll inflation.
The refined de Sitter conjecture imposes constraints on the gradient and curvature of the potential, which can be expressed in terms of the slow-roll parameters,
\begin{align}
\varepsilon = \frac{M_{\rm Pl}^2}{2}\left(\frac{V'}{V}\right)^2 > \frac{c^2}{2}
\quad\textrm{or}\quad \eta = M_{\rm Pl}^2\,\frac{V''}{V} < - c' \,,
\end{align}
where $c$ and $c'$ are two positive numbers of $\mathcal{O}\left(1\right)$.
In DHI, $\varepsilon$ is always severely suppressed, $\varepsilon \ll \eta$; at the local maximum at $s_{\rm max}$, it even vanishes.
The relevant condition in our case is therefore the constraint on $\eta$.
However, as we require our model to reproduce the best-fit value for $n_s$, we know that $\left|\eta\right|$ cannot be of $\mathcal{O}\left(1\right)$ or even larger.
Instead, we have $\eta \simeq \left(n_s^{\rm obs} - 1\right)/2 \simeq -0.01755$ at the time of CMB horizon exit.
This means that the refined de Sitter conjecture can only be satisfied if the coefficient $c'$ should turn out to be suppressed by a factor of $\mathcal{O}\left(10^{-2}\right)$ compared to the naive expectation $c' \sim 1$.
This discrepancy requires more work; however, it is important to note that it is not specific to DHI but rather concerns most models of single-field slow-roll inflation.


\smallskip\noindent\textit{Initial conditions.}
In the parameter regime considered in this paper, DHI amounts to a particular realization of hilltop inflation.
This indicates that it should be possible to satisfy the TCC bound also in other inflation models that equally feature a very flat local maximum in the potential.
In our case, the hilltop arises from the interplay of logarithmic radiative corrections and a soft SUSY-breaking mass in supergravity; but it is well conceivable that a potential similar to the one in Eq.~\eqref{eq:V} could also have a different origin at high energies.
Meanwhile, inflaton near a hilltop also turns the spotlight on the initial conditions of inflation.
In order to obtain viable inflation that ends in a waterfall phase transition, the initial inflaton field value $s_{\rm ini}$ must be located on the correct side of the hilltop, $s_{\rm ini} < s_{\rm max}$.
At face value, this suggests that $s_{\rm ini}$ must be tuned.
On the other hand, one may speculate that the inflaton field initially starts out on the other side of the hilltop, $s > s_{\rm max}$, and then jumps over or tunnels through the potential barrier.
Such a mechanism may require modifications of the potential at large field values and we leave a more detailed investigation for future work.
Alternatively, one may suppose that the GUT embedding of DHI at high energies provides a dynamical mechanism that automatically positions the inflaton field near the local maximum in the potential.


In addition to the initial inflaton field value $s_{\rm ini}$, the initial inflaton velocity $\dot{s}_{\rm ini}$ also appears to be tuned.
As pointed out in~\cite{Bedroya:2019tba}, this follows from the fact that the TCC automatically implies that $\dot{s}_{\rm ini}^2/V_{\rm ini}^{\vphantom{2}} \sim \varepsilon \ll 1$, which means that the initial kinetic energy of the inflaton field must be severely suppressed compared to its initial potential energy.
There is no straightforward reason why this should be the case.
Naively, one would rather expect the initial energy density to be more or less equilibrated, $\dot{s}_{\rm ini}^2 \sim V_{\rm ini}^{\vphantom{2}}$.
The small initial inflaton velocity at the beginning of the inflationary trajectory thus represents another open question that needs to be addressed in the GUT embedding of our model at high energies.
As before, we stress that this issue is not specific to DHI but rather a general implication of the TCC that applies to essentially all models of small-field inflation.


\newpage

\noindent\textit{Observational signatures.}
There are several ways to test the scenario proposed in this paper.
For instance, one way to falsify it would consist in the observation of a large tensor-to-scalar ratio in future CMB experiments.
An $r$ value as large as, say, $r \sim 0.001 \cdots 0.1$ would contradict the TCC bound in Eq.~\eqref{eq:rmax} and hence point to a different origin of the primordial tensor spectrum than the usual vacuum fluctuations of the metric tensor.
In this case, one would have to consider inflaton models that provide additional sources of primordial tensor perturbations, or abandon the paradigm of cosmic inflation altogether.


Another important phenomenological aspect is that our model, as defined in Eqs.~\eqref{eq:W} to \eqref{eq:VD}, results in the production of cosmic strings during the waterfall phase transition after inflation.
The tension $\mu_{\rm cs}$ of these strings is determined by the energy scale of spontaneous symmetry breaking, \textit{i.e.}, the VEV of the waterfall field $\Phi$,
\begin{align}
\label{eq:GmuCS}
G\mu_{\rm cs} \simeq 2.1\times10^{-6}\left(\frac{v}{10^{16}\,\textrm{GeV}}\right)^2 \,,
\end{align}
where $G = \left(8\pi M_{\rm Pl}^2\right)^{-1}$ denotes Newton's constant and where we used that the strings produced after DHI satisfy the Bogomolny limit, $\mu_{\rm cs} = \pi v^2$~\cite{Bogomolny:1975de}.
The string tension in Eq.~\eqref{eq:GmuCS} exceeds the current CMB bound on $G\mu_{\rm cs}$ by roughly an order of magnitude, $G\mu_{\rm cs}^{\rm cmb} \lesssim 10^{-7}$~\cite{Charnock:2016nzm,Lizarraga:2016onn}.
At first sight, this represents a problem.
However, at second sight, the production of cosmic strings after DHI could also turn into virtue; namely, if they should be unstable, which would be the case if the gauge group of our model was embedded in a semisimple GUT gauge group at higher energies~\cite{Vilenkin:1982hm} (see also~\cite{Monin:2008mp,Dror:2019syi}).
This is a reasonable assumption and could lead to a promising scenario where the string network first emits an observable signal in gravitational waves~\cite{Auclair:2019wcv} and then decays during the radiation-dominated era, so as to circumvent the CMB bound on $G\mu_{\rm cs}$.
We will come back to such a scenario in future work~\cite{Buchmuller:2019gfy}.
As an alternative to decaying cosmic strings, one may also extend our model by additional symmetry-breaking fields that are marginally coupled to the waterfall fields of DHI and that break the $U(1)$ symmetry already during inflation.
In this case, cosmic strings would be produced at early times and subsequently strongly diluted in consequence of the exponential expansion during inflation (see, \textit{e.g.},~\cite{Domcke:2017xvu,Domcke:2017rzu}).


Finally, we mention that our model has interesting implications for the particle spectrum at low energies.
The key observation is that the spontaneous $U(1)$ breaking during the waterfall transition results in a massive vector multiplet with a supersymmetric mass contribution
\begin{align}
\label{eq:mV}
m_V = gv = 100\,\textrm{GeV}\left(\frac{g}{10^{-14}}\right)\left(\frac{v}{10^{16}\,\textrm{GeV}}\right) \,.
\end{align}
This multiplet consists of the real $U(1)$ Higgs boson, the $U(1)$ Higgsino, the $U(1)$ gaugino, and the massive $U(1)$ vector boson.
At the end of the day, the masses of these particles are potentially vastly separated from each other, depending on the details of SUSY breaking and its mediation to the waterfall sector.
It is still an interesting observation that the combination of the GUT-scale VEV $v$ and the tiny gauge coupling constant $g$ readily results in a vector mass in the vicinity of the electroweak scale.
Therefore, depending on the details of the coupling between the waterfall fields and the standard model, some of the particles contained in the massive $U(1)$ vector multiplet could be within the reach of collider experiments.
More work in this direction is certainly desirable.


\smallskip\noindent\textit{Reheating temperature.}
The coupling between the inflaton sector and the standard model also determines the reheating temperature after inflation.
In the following, we will not specify the exact nature of this coupling, which requires further model building and is thus beyond the scope of this work.
Instead, we will merely assume that the duration of reheating is controlled by the perturbative decay of the $U(1)$ Higgs field $\phi$ (see also \cite{Buchmuller:2010yy,Buchmuller:2011mw,Buchmuller:2012wn,Buchmuller:2012bt,Buchmuller:2013lra,Schmitz:2012kaa,Buchmuller:2013dja,Domcke:2017xvu,Domcke:2017rzu}).
Then, if we parametrize the $\phi$ decay rate $\Gamma_\phi$ in a model-independent way by an effective coupling constant $\lambda_{\rm eff}$, 
\begin{align}
\Gamma_{\phi} = \frac{\lambda_{\rm eff}^2}{8\pi}\,m_\phi \,,
\end{align}
the reheating temperature can be estimated as follows,
\begin{align}
\label{eq:Treh}
T_{\rm reh} & = \left(\frac{90}{\pi^2g_{*,\rho}}\right)^{1/4}\sqrt{\Gamma_\phi\,M_{\rm Pl}}
\\ \nonumber
& \simeq 140\,\textrm{TeV}\left(\frac{\lambda_{\rm eff}}{10^{-4}}\right)\bigg(\frac{m_\phi}{100\,\textrm{GeV}}\bigg)^{1/2} \,.
\end{align}
Here, we have chosen a scalar mass $m_\phi$ of the same order of magnitude as the vector mass in Eq.~\eqref{eq:mV} and an effective coupling constant $\lambda_{\rm eff}$ that allows us to avoid the thermal overproduction of gravitinos (see below).


At temperatures of $\mathcal{O}\left(100\right)\,\textrm{TeV}$, the baryon asymmetry of the Universe can be produced via thermal leptogenesis~\cite{Fukugita:1986hr}, provided there is a slight degeneracy in the heavy-neutrino mass spectrum that resonantly enhances the $CP$ asymmetry in heavy-neutrino decays~\cite{Pilaftsis:1997dr,Pilaftsis:1997jf,Pilaftsis:2003gt,Dev:2017wwc} (see also~\cite{Brdar:2019iem,Brivio:2019hrj} for recent work on resonant leptogenesis).
Alternatively, there is a variety of leptogenesis scenarios in the literature that are based on minimal extensions of the simplest type-I seesaw model and that accomplish to successfully generate the baryon asymmetry at temperatures $T \lesssim 100\,\textrm{TeV}$ (see \cite{Hugle:2018qbw,Borah:2018rca,Alanne:2018brf} for three recent examples based on scalar extensions of the type-I seesaw model).


\smallskip\noindent\textit{Weak-gravity conjecture.}
Our inflation model is also consistent with the weak-gravity conjecture (WGC)~\cite{ArkaniHamed:2006dz}.
The electric WGC requires the presence of states with charge $q$ and mass $m < m_{\rm WGC} = \sqrt{2}\,gq\,M_{\rm Pl}$ in the particle spectrum.
This requirement is readily satisfied by the particles in the massive vector multiplet, whose mass is induced by the GUT scale rather than the Planck scale,  $m_V/m_{\rm WGC} \sim v / M_{\rm Pl} \ll 1$ [see Eq.~\eqref{eq:mV}].
The magnetic WGC meanwhile states that the effective field theory describing the dynamics of the inflaton field is only valid up to a UV cutoff scale $\Lambda_{\rm WGC} \sim g M_{\rm Pl}$, which is at most of $\mathcal{O}\left(100\right)\,\textrm{TeV}$ in our scenario.
The scale $\Lambda_{\rm WGC}$ needs to be compared to the inflaton-dependent energy scales in the scalar potential:
the inflaton field value, which is bounded by $s_{\rm max} \lesssim 100\,\textrm{TeV}$ [see Eq.~\eqref{eq:smax}], and the supergravity and radiative terms, whose fourths roots do not become larger than $\mathcal{O}\left(100\right)\,\textrm{GeV}$.
All relevant interactions of the inflaton field thus occur at energies below the UV cutoff scale $\Lambda_{\rm WGC}$.
By contrast, the energy density stored in the vacuum, $V_D^0 = 1/2\,g^2\xi^2$, is assumed to originate from a separate sector (see footnote~\ref{fn:xi}).
This GUT sector is responsible for GUT symmetry breaking, the generation of the FI term $\xi$, and necessarily requires a larger UV cutoff scale.%
\footnote{The inflaton field is obviously not responsible for GUT symmetry breaking.
It is thus not necessary to have a reliable description of the inflaton dynamics all the way up to the GUT scale.}
In this paper, we shall not specify the GUT embedding of our model.
For our purposes, it suffices to assume that the GUT sector contains a set of appropriately charged moduli whose VEVs induce a $U(1)$ FI term in the inflaton sector.
Given the fact that the inflaton is a $U(1)$ singlet, it is reasonable to assume that this kind of communication between the GUT and inflaton sectors has no further unwanted consequences for the dynamics of inflation.
A more detailed investigation of possible GUT embeddings is left for future work.


\smallskip\noindent\textit{Gravitino dark matter.}
A gravitino mass of $\mathcal{O}\left(10\right)\,\textrm{MeV}$ [see Eq.~\eqref{eq:m32g}] in combination with a reheating temperature of $\mathcal{O}\left(100\right)\,\textrm{TeV}$ [see Eq.~\eqref{eq:Treh}] are the perfect basis for DM in the form of thermally produced gravitinos.
To leading order in the strong gauge coupling constant, supersymmetric QCD results in the following present-day abundance of thermally produced gravitinos~\cite{Bolz:2000fu,Pradler:2006qh},
\begin{align}
\label{eq:h2O32}
h^2\Omega_{3/2} \simeq 0.39\left(\frac{T_{\rm reh}}{100\,\textrm{TeV}}\right)\left(\frac{10\,\textrm{MeV}}{m_{3/2}}\right)\bigg(\frac{m_{\tilde{g}}}{3\,\textrm{TeV}}\bigg)^2 ,
\end{align}
where $m_{\tilde{g}}$ denotes the gluino mass at energies around the $Z$-boson mass.
Including additional processes in the MSSM coupled to gravity, the total gravitino abundance can still be enhanced by an additional factor of $\mathcal{O}\left(1\right)$~\cite{Rychkov:2007uq}; in the following, we will, however, content ourselves with the simple estimate in Eq.~\eqref{eq:h2O32}.
Based on this estimate, we can now solve the condition that gravitinos should account for the entire DM relic density, $h^2\Omega_{3/2} = h^2\Omega_{\rm dm} \simeq 0.12$~\cite{Aghanim:2018eyx,Akrami:2018odb}, for the gluino mass $m_{\tilde{g}}$,
\begin{align}
m_{\tilde{g}} \simeq  1.7\,\textrm{TeV} \bigg(\frac{m_{3/2}}{10\,\textrm{MeV}}\bigg)^{1/2}\left(\frac{100\,\textrm{TeV}}{T_{\rm reh}}\right)^{1/2} \,.
\end{align}
For $T_{\rm reh} = 100\,\textrm{TeV}$, the upper bound on $m_{3/2}$ in Eq.~\eqref{eq:m32max} thus results in an upper bound on the gluino mass of $m_{\tilde{g}}^{\rm max} \simeq 2.5\,\textrm{TeV}$.
This defines an interesting gluino mass range and opens the exciting possibility to probe the scenario of gravitino DM in future collider experiments.


In closing, we mention that a similar mass range for gravitino DM as in our scenario has recently been identified in~\cite{Choi:2019jck}, however, for a completely different reason.
The starting point of~\cite{Choi:2019jck} is the extension of the MSSM by a supersymmetric axion multiplet that couples to the electroweak gauge bosons.
This electroweak axion receives an extremely flat potential from $SU(2)_L$ instantons, which turns it into a quintessence field that serves as an explanation of dark energy~\cite{Nomura:2000yk,Ibe:2018ffn}.
The authors of~\cite{Choi:2019jck} consider the decay of unstable gravitino dark matter into the degrees of freedom contained in the electroweak axion multiplet, $\tilde{G}\rightarrow a + \tilde{a}$, where $a$ denotes the axion quintessence field and $\tilde{a}$ is the corresponding axino.
They argue that this decay represents a concrete realization of the decaying-DM mechanism proposed in~\cite{Vattis:2019efj} to resolve the so-called Hubble tension, \textit{i.e.}, to reconcile the CMB measurement of the Hubble parameter $H_0$ by the PLANCK satellite with the local $H_0$ value obtained via the cosmic distance ladder and time delay measurements of strongly lensed quasars~\cite{Wong:2019kwg}.
The main idea behind this solution to the Hubble tension is to slowly transfer energy from cold dark matter to massless radiation (\textit{i.e.}, axions in~\cite{Choi:2019jck}) and a subdominant component of warm dark matter (\textit{i.e.}, axinos in~\cite{Choi:2019jck}).
The energy in radiation is then quickly redshifted away, which moves the onset of the dark-energy-dominated era to earlier times. 
This in turn leads to a slower decrease of the Hubble rate in the late Universe, which eventually results in a larger $H_0$ value at the present time than what one would expect in the standard $\Lambda$-cold-dark-matter ($\Lambda$CDM) picture.


A necessary condition for this mechanism to work is that the lifetime of the decaying DM particle, $\tau$, falls in the range $1.3\lesssim \log_{10}\left(\tau/\textrm{Gyr}\right) \lesssim 2.2$~\cite{Vattis:2019efj}.
The authors of~\cite{Choi:2019jck} use this and other constraints on the model to derive a viable mass range for decaying gravitino dark matter, $m_{3/2} \sim 100\,\textrm{MeV}\cdots 1\,\textrm{GeV}$.
This differs from our gravitino mass range only by a factor $\mathcal{O}\left(10\cdots100\right)$.
In our case, the gravitino is unfortunately too long-lived.
Neglecting any phase space factors, we can estimate the gravitino lifetime due to the decay $\tilde{G}\rightarrow a + \tilde{a}$ as follows, 
\begin{align}
\tau \sim \frac{192\pi M_{\rm Pl}^2}{m_{3/2}^2} \sim 6000\,\textrm{Gyr}\left(\frac{23\,\textrm{MeV}}{m_{3/2}}\right)^3 \,,
\end{align}
which exceeds the upper bound on $\tau$ derived in~\cite{Vattis:2019efj} by about a factor $40$.
Nonetheless, it is an remarkable coincidence that the TCC bound on DHI points to a mass range for gravitino DM that is close to the one considered in~\cite{Choi:2019jck}.
We believe that the possible connection between the TCC, DHI, axion quintessence, and a resolution of the Hubble tension due to decaying gravitino DM deserves more attention in future work.
It would be interesting to consider, \textit{e.g.}, scenarios with a relaxed upper bound on the gravitino mass in consequence of nonstandard dynamics during reheating, $-1/3 < w <0$~\cite{Mizuno:2019bxy}.


\medskip\noindent\textbf{Conclusions.} 
In this paper, we discussed the implications of the recently proposed trans-Planckian censorship conjecture (TCC) for single-field slow-roll inflation and showed how the resulting constraint on the inflationary energy scale can be readily satisfied in a minimal D-term hybrid inflation (DHI) model.
Our main observation was that a tiny gauge coupling constant $g$ always allows us to parametrically suppress the energy scale of inflaton compared to the GUT scale $\sqrt{\xi} \sim 10^{16}\,\textrm{GeV}$.
In future work, it would be interesting to construct an explicit GUT embedding of our model that sheds more light on the question where such a small value of $g$ might come from.


We also discussed: the relation between our model and other quantum gravity swampland conjectures; several issues related to the initial conditions of inflation; the production of cosmic strings; the prospects to probe our scenario in CMB observations, in gravitational-wave experiments, and at colliders; the thermal generation of gravitino dark matter; and the possibility to relate our model to scenarios that explain dark energy in terms of axion quintessence and that resolve the Hubble tension by means of decaying gravitino dark matter.
In this discussion, we identified masses of $\mathcal{O}\left(10\right)\,\textrm{MeV}$ as the relevant mass range for gravitino dark matter and masses of $\mathcal{O}\left(100\right)\,\textrm{GeV}$ as the mass range of the degrees of freedom in the waterfall sector that are responsible for reheating (and which should thus couple to the standard model). 
In summary, we conclude that DHI based on a tiny gauge coupling constant results in an interesting theoretical framework as well as in an exciting phenomenology.
Both aspects of our model await further exploration.


\bigskip\noindent\textit{Acknowledgements.} I wish to thank Wilfried Buchm\"uller, Valerie Domcke, and Hitoshi Murayama for discussions on unstable cosmic strings and the related signal in gravitational waves; and I wish to thank Tsutomu Yanagida for comments on the relation between the model studied in this paper and the weak-gravity conjecture.
This project has received funding from the European Union's Horizon 2020 Research and Innovation Programme under grant agreement number 796961, ``AxiBAU''.


\bibliographystyle{JHEP}
\bibliography{arxiv_2}


\end{document}